\documentclass[aps,preprint,showpacs]{revtex4}
\usepackage[dvips]{graphicx}
\usepackage{amsmath}
\usepackage{color}

\begin{document}
\draft

\title{\bf Experimental and numerical study of spectral properties of three-dimensional chaotic microwave cavities: The case of missing levels}

\author{Vitalii Yunko, Ma{\l}gorzata Bia{\l}ous, Szymon Bauch, Micha{\l} {\L}awniczak, and Leszek Sirko}

\address{Institute of Physics, Polish Academy of Sciences, Aleja  Lotnik\'{o}w 32/46, 02-668 Warsaw, Poland}

\date{\today}

\bigskip

\begin{abstract}

We present an experimental and numerical study of missing-level statistics of chaotic three-dimensional microwave cavities. The nearest-neighbor spacing distribution, the spectral rigidity, and the power spectrum of level fluctuations were investigated. We show that the theoretical approach to a problem of incomplete spectra does not work well when the incompleteness of the spectra is caused by unresolved resonances.  In such a case the fraction of missing levels can be evaluated by calculations based on random matrix theory.

\end{abstract}

\bigskip
\maketitle
\section{Introduction}

Low-dimensional microwave systems are extremely useful for studying quantum chaos. One-dimensional (1D) microwave networks \cite{Hul2004,Lawniczak2010,Hul2012} can be used to simulate quantum graphs due to the equivalence of the telegraph equation describing them and the Schrodinger equation describing corresponding quantum graphs \cite{Hul2004,Kottos1997,Kottos1999,Pakonski2001,Sirko2016}. In turn, two-dimensional (2D) microwave cavities  \cite{Stoeckmann90,Sridhar,Richter,So95,Stoffregen1995,Sirko1997,Hemmady2005,Dietz2015} can be used to simulate quantum billiards due to the analogy of the Helmholtz scalar equation and the Schr\"odinger equation describing these systems, respectively. In the case of three-dimensional (3D) microwave cavities, there is no direct analogy between Helmholtz's vector equation and the Schr\"odinger equation. Thus, the cavities cannot simulate quantum systems. Nevertheless, the spectral statistics of irregular/rough 3D cavities display behavior characteristic for classically chaotic quantum systems  \cite{Sirko1995,Alt1997,Dorr1998,Dembowski2002,Tymoshchuk2007,Savytskyy2008,Schroeder1954,Lawniczak2009,Yeh2013}. Therefore, the 3D microwave cavities are also very interesting objects in the research of the properties of wave chaos. However, studies of such systems, both theoretical, including properties of random electromagnetic vector fields \cite{Primack2000,Prosen1997,Arnaut2006,Gros2014}, as well as experimental are very rare.

In the spectral measurements of any systems the loss of some resonances is inevitable. It may be caused by low signal-to- noise ratio, by degeneration or overlap of resonances due to losses (absorption/system openness). In the case of 3D systems we have an additional obstacle: a large density of states. However, one should mention that in the studies of spectral statistics of acoustic resonances in 3D aluminum  \cite{Weaver1989} and quartz \cite{Ellegaard1996} blocks, characterized by high quality factors $Q\simeq 10^4-10^5$, no missing resonances were reported. In billiard systems, even higher quality factors were obtained in the experiments with superconducting microwave
cavities \cite{Alt1997}. In normal conducting resonators
the quality factors are much lower ($Q \simeq 10^3$) and the loss of
some modes is either very likely or even inevitable, therefore,  in the recent study of such a chaotic 3D microwave cavity the missing levels were explicitly taken into account \cite{Lawni2018a}.

The determination of the system chaoticity and symmetry class defined in the random matrix theory (RMT) using its spectral properties requires knowledge of complete series of eigenvalues, so experimentally it is generally a difficult task  \cite{Bohigas1984,Bohigas1983,Sieber1993,Dietz2014}. The procedures developed for microwave networks with broken  \cite{Bialous2016} and preserved  \cite{Dietz2017} time reversal symmetries (TRS) show that this is possible provided that several statistical measures, e.g. a short-range correlation function (the nearest-neighbor spacing distribution - NNSD), long-range correlation functions (e.g., the spectral rigidity), and the power spectrum of level fluctuations  \cite{Molina2007,Relano2002,Faleiro2004} will be analyzed. Nuclei and molecules  \cite{Frisch2014,Mur2015,Liou1972,Zimmermann1988} are examples of the real physical systems for which such procedures are crucial because in their case one always deals with incomplete spectra \cite{Enders2000,Agvaanluvsan2003,Bohigas2004}.

\section{Experimental setup and measurements}

The overall view of the experimental setup is shown in Fig.~1(a). The 3D microwave cavity was made of polished aluminum type EN 5754 and consists of four elements. The rough semicircular element of height 60 mm (marked by (1) in Fig.~2 and visible in Fig. 1(b)) is closed by two flat parts: side (labeled by (2) in Fig.~2) and upper ones, both visible in Fig.~2. The bottom element is a slightly inclined and convex plate (labeled by (3) in Fig.~2) preventing the appearance of bouncing balls orbits between the upper and lower walls of the cavity. The radius function $R(\theta) = R_0 +\sum^M _{m=2} a_m
\sin(m\theta+\Phi_m)$, where the mean radius $R_0$ =10.0 cm, $M$ = 20, $a_m$ and $\Phi_m$ are uniformly distributed on [0.084, 0.091] cm and on [0, 2$\pi$], and 0$\leq\theta<\pi$, described the rough, semicircular element on the plane of the cross-section (Fig.~2). An aluminum scatterer inside the cavity, mounted on the metallic axel in its upper wall (see Fig.~1) was used to realize various cavity configurations. The orientation of the scatterer was changed by turning the axle around in $18$ identical steps, each equal to $\pi /9$. Besides the hole for the scatterer in the bottom wall there are three other holes A1, A2, and A3 for antennas.

An Agilent E8364B vector network analyzer (VNA) was used to perform two-port measurements of the four-element scattering matrix $\hat S$  \cite{Lawniczak2015,Lawni2018a}.

\begin{equation}
\label{Smatrix}
\hat S=\left[
\begin{array}{c c}
S_{11} & S_{12}\\
S_{21} & S_{22}
\end{array}
\right]
\end{equation}

The cavity was connected to the VNA via two antennas and the flexible microwave cables HP 85133-616 and HP 85133-617.  The measurements were done in the frequency range 6-11 GHz for all three combinations of the antennas positions. The antennas penetrated 6 mm into the cavity with a wire of 0.9 mm in diameter. The "third" empty hole was plugged by brass plug during the measurement. Previous measurements \cite{Lawniczak2009} have shown that the total absorption of the cavity is mainly related to internal absorption much greater than that associated with antennas/channels.

In Fig.~3 we present the examples of the modules of the reflected $|S_{11}|$, $|S_{22}|$, and transmitted $|S_{12}|$ signals measured in the lowest $6-7$ GHz and highest $10-11$ GHz frequency ranges. It should be noted that a full cross-correlation was observed between $S_{12}$ and $S_{21}$, which means that TRS is preserved.   The spectra were obtained in the two-port measurement with the antennas at positions 2 and 3 (Fig.~1(c)) connected to the ports 1 and 2 of the VNA, respectively. In the low frequency range, when the resonances are well separated, comparing the spectra $|S_{11}|$ and $|S_{22}|$, it is clearly seen that the number of detected resonances may depend on the position of the antenna. The transmission signal $|S_{12}|$ can be also used in the search for the resonances, however, even for this low frequency range not all resonances are separated and visible. In turn, in the higher frequency range, due to the cubic dependence of the number of resonances on the frequency, the overlapping resonances have appeared. Therefore, losing resonances in the measurements is inevitable.

\section{Statistical measures of experimental spectra of chaotic systems}

In order to analyze experimental data, it is useful to eliminate their dependence on specific features of the studied system, such as its dimensions, for example. This may be achieved by the rescaling procedure, which in the case of the 3D chaotic systems is carried out using the Weyl formula \cite{Balian,Balian1971,Baltes1972,Gros2014}:

\begin{equation}
N(\nu)=A\nu^3-B\nu + C.
\label{Bloch}
\end{equation}

The coefficient $A=\frac{8}{3}\frac{\pi}{c^3}V$, where $c$ is speed of light in vacuum and $V = (7.267\pm 0.012)\times 10^{-4}$ m$^3$ is the volume of the cavity reduced by the volume of the scatterer. The dependent on the surface of a cavity, a term proportional to $\nu^{2}$ disappears due to boundary conditions of the electromagnetic field in the conducting cavity walls \cite{Balian1971}. The coefficient $B$ depends on the surface curvature, internal angles, and the edge length of the cavity \cite{Gros2014}. The constant C is also associated with the shape of the cavity and in the simple case of the cubic cavity $C=1/2$ \cite{Baltes1972}. The coefficients $B$ and $C$ are generally difficult to determine exactly, so fitting procedures are necessary to obtain the cumulative number of levels $N(\nu)$ for 3D irregular cavities.

In order to analyze the data we will use the short-range spectral fluctuation function, the nearest-neighbor spacing distribution, i.e. the distribution of spacings between adjacent eigenvalues $s_i=\epsilon_{i+1}-\epsilon_i$, where $\epsilon_i=N\left(\nu_i\right)$ are rescaled eigenvalues obtained by unfolding procedure will be used. Also, the integrated nearest-neighbor spacing distribution $I(s)$, very sensitive to the symmetry class of the system (preserved or broken time reversal symmetry) will be used. The spectral rigidity of the spectrum $\Delta_3(L)$ which is the least-squares deviation of the integrated resonance density from the straight line that best fits it in the interval $L$ \cite{Mehta1990} will be exploit as the measure of long-range spectral fluctuations. We also will take into account the power spectrum of the deviation of the $q$th nearest-neighbor spacing from its mean value $q$ \cite{Molina2015}.

\begin{equation}
\eta_q= \sum_{i=1}^{q}(s_{i}-<s>)=\epsilon_{q+1}-\epsilon_1-q
\label{Blochb}
\end{equation}

In the systems with losses, when the problem of missing levels may be very severe, the fluctuations of the scattering matrix elements can be useful. The correlation functions \cite{Dietz2009,Dietz2010}, the Wigner's reaction matrix and the elastic enhancement factor \cite{Lawniczak2010,Sirko2016,Lawniczak2015} are sensitive measures of system chaoticity, however don't provide any information about missing energy levels.

In the article \cite{Bohigas2004} the authors derived analytical expressions for the spectral rigidity, the number variance, and the nearest-neighbor spacing distribution describing incomplete spectra. The parameter $0<\varphi \leq1$ denotes the fraction of observed levels. The formula for the spectral rigidity reads as follows:

\begin{equation}
\delta_3(L)=(1-\varphi)\frac{L}{15} + \varphi^2\Delta_3\left(\frac{L}{\varphi}\right).
\label{delta3}
\end{equation}

where $\Delta_3(L)$ is the expression for the complete spectra.

The NNSD is expressed by the sum of terms of the $(n+1)$st nearest-neighbor spacing distribution $P(n,\frac{s}{\varphi})$:

\begin{equation}
p(s)= \sum_{n=0}^{\infty}(1-\varphi)^{n}P(n,\frac{s}{\varphi}).
\label{pdistr}
\end{equation}

For GOE systems the first and second term of Eq. (5) is approximated by

\begin{equation}
P(0,\frac{s}{\varphi}) = \frac{\pi}{2}\frac{s}{\varphi}\exp\left[-\frac{\pi}{4}\left(\frac{s}{\varphi}\right)^{2}\right].
\label{p0distr}
\end{equation}

\begin{equation}
P(1, \frac{s}{\varphi})=\frac{8}{3\pi^{3}}\left(\frac{4}{3}\right)^5\left(\frac{s}{\varphi}\right)^4\exp\left[-\frac{16}{9\pi}\left(\frac{s}{\varphi}\right)^2\right].
\label{p1distr}
\end{equation}

When $\varphi=1$ the above formulas reduces to Wigner surmise formula for the NNSD for $P(0,s)$ and to the NNSD of the symplectic ensemble with $\left<s\right>=2$ for $P(1,s)$.
For the higher $n=2,3,\ldots$ spacing distributions $P(n, \frac{s}{\varphi})$, are well approximated by their Gaussian asymptotic forms, centered at $n+1$.

\begin{equation}
P(n,\frac{s}{\varphi})= \frac{1}{\sqrt{2\pi V^2(n)}}\exp\left[-\frac{(\frac{s}{\varphi}-n-1)^2}{2V^2(n)}\right],
\label{pndistr}
\end{equation}

with the variances

\begin{equation}
V^2(n) \simeq \Sigma^2(L=n)-\frac{1}{6}.
\label{vvariance}
\end{equation}

The number variance $\Sigma^2(L)$ in Eq.~(\ref{vvariance}) is the variance of the number of levels contained in an interval of length $L$ \cite{Mehta1990}.

The integrated nearest-neighbor spacing distribution $I(s)$, which is very useful for distinguishing of GOE and GUE (Gaussian unitary ensemble - systems with broken TRS) due to its high sensitivity from s in the range of small level separations, reads:

\begin{equation}
I(s)= \int^s_0p(s')ds'.
\label{ipdistr}
\end{equation}

The analytical expression for the power spectrum of level fluctuations for incomplete spectra was given in Ref. \cite{Molina2007}

\begin{eqnarray}
\langle S(\tilde k)\rangle &=&\nonumber
\frac{\varphi}{4\pi^2}\left[\frac{K\left(\varphi\tilde k\right)-1}{\tilde k^2}+\frac{K\left(\varphi\left(1-\tilde k\right)\right)-1}{(1-\tilde k)^2}\right]\\
&+& \frac{1}{4\sin^2(\pi\tilde k)} -\frac{\varphi^2}{12},
\label{noise}
\end{eqnarray}

where $K(\tau)=2\tau-\tau\log(1+2\tau)$ is the spectral form factor for GOE system and $0\leq\widetilde{k} = k/N\leq1$.

The $<S(k)>$ is given in terms of the Fourier spectrum transform from "time" $q$ to $k$

\begin{equation}
S(k)=|\tilde{\eta}_k |^2
\label{skform}
\end{equation}
with
\begin{equation}
\tilde{\eta}_k=\frac{1}{\sqrt{N}}\sum_{q=0}^{N-1} \eta_q\exp\left(-\frac{2\pi ikq}{N}\right)
\label{delta}
\end{equation}

In Refs. \cite{Relano2002,Faleiro2004} authors showed that for $\tilde k \ll 1$ the  $\langle S(\tilde k)\rangle\propto (\tilde k)^{-\alpha}$ with $\alpha$ equals 2 and 1 for regular and chaotic system, respectively, regardless the system time symmetry. The power spectrum and the power law behavior were studied numerically in \cite{Robnik2005,Salasnich2005,Santhanam2005,Relano2008}. The usefulness of the power spectrum in analyzing experimental results was confirmed in \cite{Mur2015} - measurements of molecular resonances and in the investigations of microwave networks \cite{Bialous2016,Dietz2017,Lawni2017a}, and billiards \cite{Faleiro2006}.

\section{Results}

In Fig.~4 we present the experimental results obtained in the frequency range $6-11$ GHz for 30 realizations of the cavity. The number of the detected resonances depends on the cavity realization, therefore, from each cavity spectrum a small number of resonances has been randomly removed to obtain the same number $\Delta N_{exp}=208$ for all cavity configurations. The NNSD, the integrated NNSD, the spectral rigidity and the power spectrum are shown in panels (a), (b), (c) and (d), respectively. The solid (black) line denotes the theoretical predictions for the complete spectra $\varphi=1$ of the GOE system. Ten terms of Eq. (5) was used in the numerical calculation of the NNSD.  Experimental data are represented by a blue histogram and blue full diamonds in panels (a), (b) and blue full triangles and circles in panels (c), (d). Using the equations for the spectral rigidity (4) and the power spectrum (11) we found that the best agreement between the theoretical predictions for incomplete spectra (dark cyan dash-dot line) and experimental results occurs for $\varphi= 0.85$, as shown in panels (c) and (d) of Fig. 4. In turn, $\varphi=0.85$ was inserted into equations (5) and (10) to calculate the NNSD and the INNSD which are shown in panels (a) and (b). It is easily seen that the spectral rigidity and the power spectrum are much more sensitive measures of losing states than the nearest-neighbor spacing distribution. The excellent agreement between the theoretical predictions for incomplete spectra and the experimental results shows that the investigated system, 3D irregular cavity, belongs to systems characterized by preserved TRS (GOE) and that the resonances have been randomly lost.

In order to estimate the theoretical number of resonances  $\Delta N_{w}$ in the frequency range $6-11$ GHz, which is required for the calculation of  $\varphi $, we found for nine configurations of the cavity complete spectra in the frequency range 7-9 GHz \cite{Lawni2018a}.  This  allowed us to make the fits of the experimental staircase functions to the formula (2) $N(\nu)=A\nu^3-B\nu + C$
 with the fixed coefficient $A=\frac{8}{3}\frac{\pi}{c^3}V = (0.2259\pm 0.004)\times 10^{-27}s^{3}$. The fits gave the average values of the coefficients $B=(1.442 \pm 0.174)\times 10^{-9}$s and $C=-66.0 \pm 1.4$. Then using the formula (2) we calculated  the  theoretical number of resonances $\Delta N_{w} = 245$ in the frequency range $6-11$ GHz.

The fraction of the detected levels $\varphi=0.85$ estimated from the missing-level statistics (Eqs.~4 and 11) can now be compared with the fraction  $\Delta N_{exp}/\Delta N_{w} =208/245 \simeq 0.849$ obtained as a ratio of the experimentally founded eigenfrequencies and those predicted from the Weyl formula in the frequency range $6-11$ GHz. The agreement is excellent.

The discussed above theoretical and experimental results are additionally compared to the numerical results obtained directly by the application of the random matrix theory. We created 99 realizations of random, real, symmetric matrices of a size N=295, representing GOE system. 20 eigenvalues of the matrix were removed from the beginning and the end of each set of eigenvalues, yielding 245 eigenvalues as for the complete spectrum. Then, $15\%$ of the eigenvalues were randomly removed, so we finally got 208 eigenvalues as in the experiment, which were rescaled using a fifth order polynomial. The results for all considered measures are also presented in Fig. 4. The NNSD is marked by the red dotted histogram, the integrated NNSD by empty red diamonds, the spectral rigidity by empty red triangles and the power spectrum by red crosses. Again, the agreement with the experimental results is remarkable, confirming that the investigated system, 3D irregular cavity, belongs to the systems with preserved time reversal symmetry and that the resonances have been randomly lost.

From the experimental point of view it is important to analyze a more complicated situation when some of the resonances are not randomly lost. We modified the experimental spectra, originally with 15\%  of randomly missing levels, using the following procedure. In the step by step procedure in an analyzed spectrum we identified the pair of resonances, the least distant from each other, and eliminated one of them until we reached $\varphi = 0.65$.  In this way we additionally removed 20\% resonances due to clustering. In Fig. 5 we compare the results for the modified experimental spectra with the GOE prediction for the complete spectra, predictions for the missing-level statistics (Eqs. (4), (5), (10), and (11)) calculated for $\varphi = 0.65$,  and with the results of RMT calculations. It is clearly seen that the missing-level statistics (purple dashed line in all panels) fail in description of the spectra in which resonances were not lost randomly (green full line histogram - (a), green full diamonds - (b), green full triangles - (c), green dots - (d)). This is a very important problem because the loss of energy levels  due to their degeneration or overlap caused by absorption or openness of a system is very common.

The results of RMT calculation (vine doted histogram - (a), vine empty diamonds - (b), vine empty triangles - (c), vine crosses - (d))  in which the procedure of eliminating eigenvalues mimic the procedure used for the experimental data are in good agreement with experimental ones.  It should be pointed out that these numerical calculation are sensitive to the order in which the eigenvalues are deleted. The results may slightly vary depending whether we start the deletion from the overlapping resonances or from random ones.

\section{Conclusions}

We present an experimental and numerical study of the fluctuation properties in incomplete spectra of the 3D chaotic microwave cavity. We analyzed the two important cases: the situation of randomly lost resonances and the situation when an additional fraction of resonances is omitted due to their clustering (overlapping). In the case of randomly missing resonances our results are in agreement with the level-missing statistics. However, in the case of many overlapping resonances direct random matrix theory calculations are required to properly simulate the experimental results.

\section{Acknowledgement}

This work was supported in part by the National Science Centre Grants Nos. UMO-2016/23/B/ST2/03979 and 2017/01/X/ST2/00734.

\pagebreak


\smallskip

\begin{figure}[t!]
\includegraphics[width=0.5\linewidth]{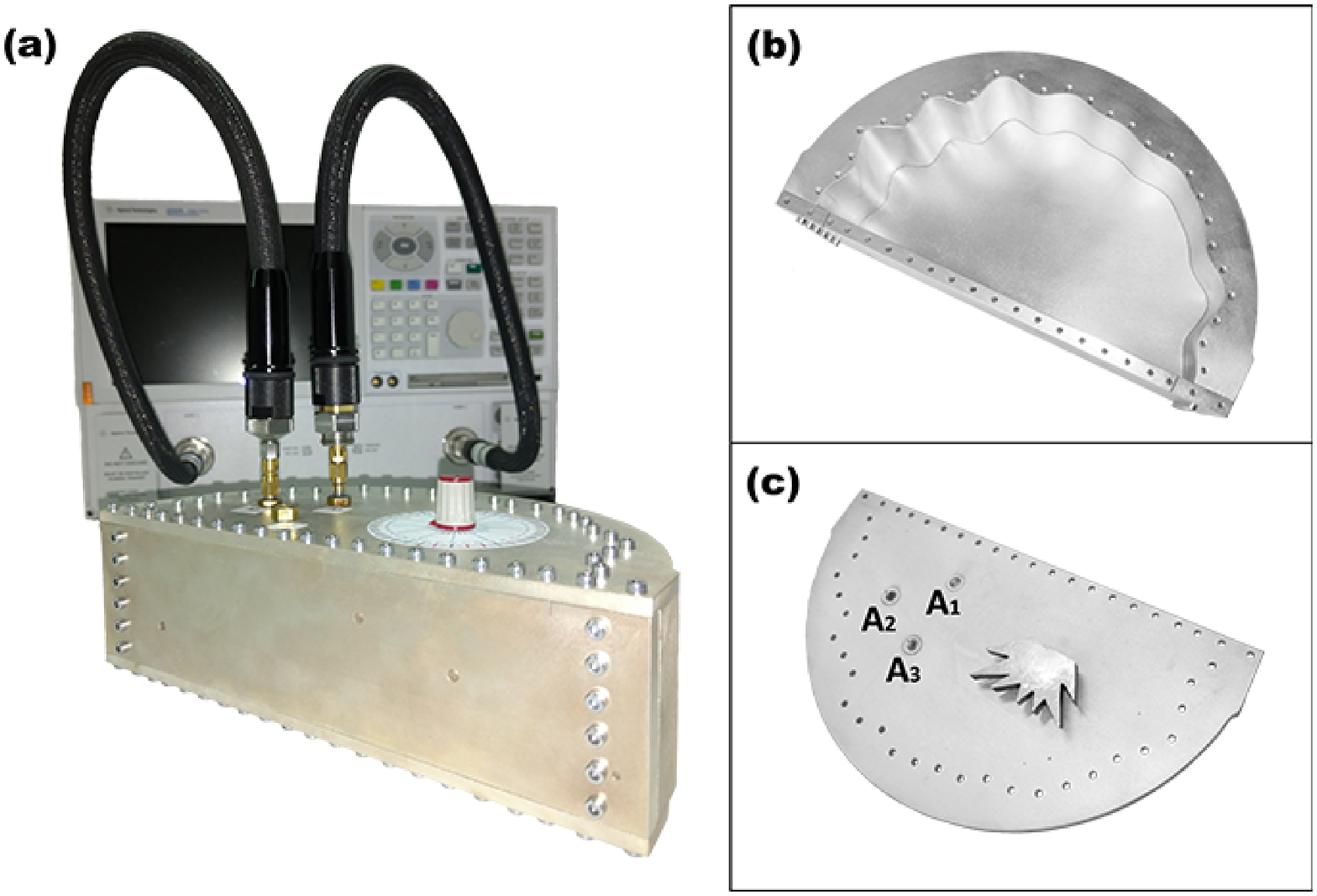}
\caption{
(a) The 3D microwave cavity connected to the vector network analyzer. The panels (b) and (c) show the cavity without the upper cover and the inner side of the upper cover with the scatterer and marked holes for the antennas.
}
\label{Fig1}
\end{figure}

\begin{figure}[h!]
\includegraphics[width=0.5\linewidth]{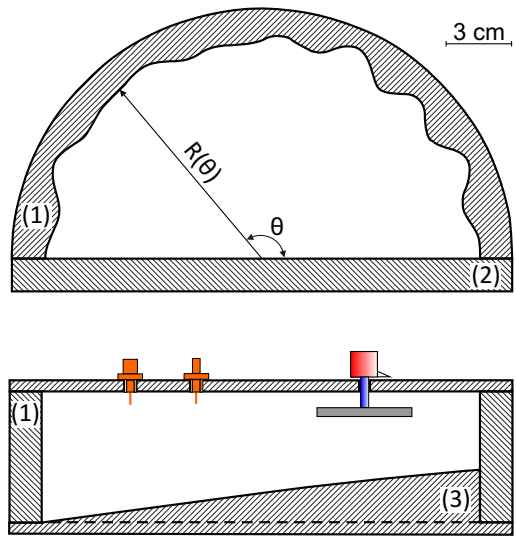}
\caption{
The sketch of the cavity. The perpendicular and parallel cross-sections to the cavity height are shown.  The positions of the scatterer and the antennas are marked. See the text for a detailed description.}
\label{Fig2}
\end{figure}

\begin{figure}[h!]
\includegraphics[width=0.5\linewidth]{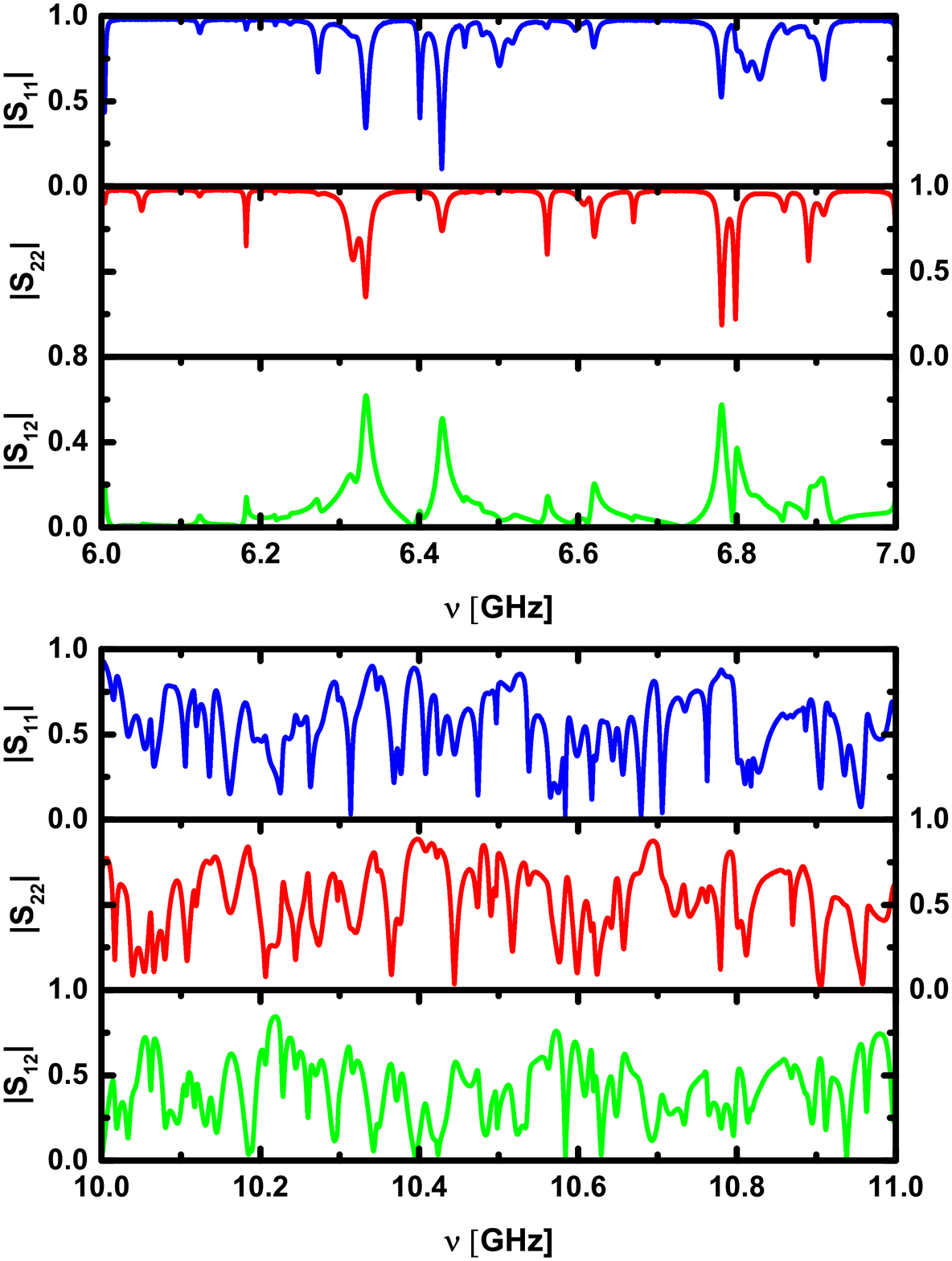}
\caption{
Examples of the measured modules of the elements $|S_{11}|$ $|S_{22}|$ and $|S_{12}|$ of the scattering matrix $\hat S$ (antennas in holes A$_2$ and A$_3$) of the 3D microwave cavity in the frequency range $6-7$ GHz and $10-11$ GHz, respectively.
}
\label{Fig3}
\end{figure}

\begin{figure}[h!]
\includegraphics[width=0.9\linewidth]{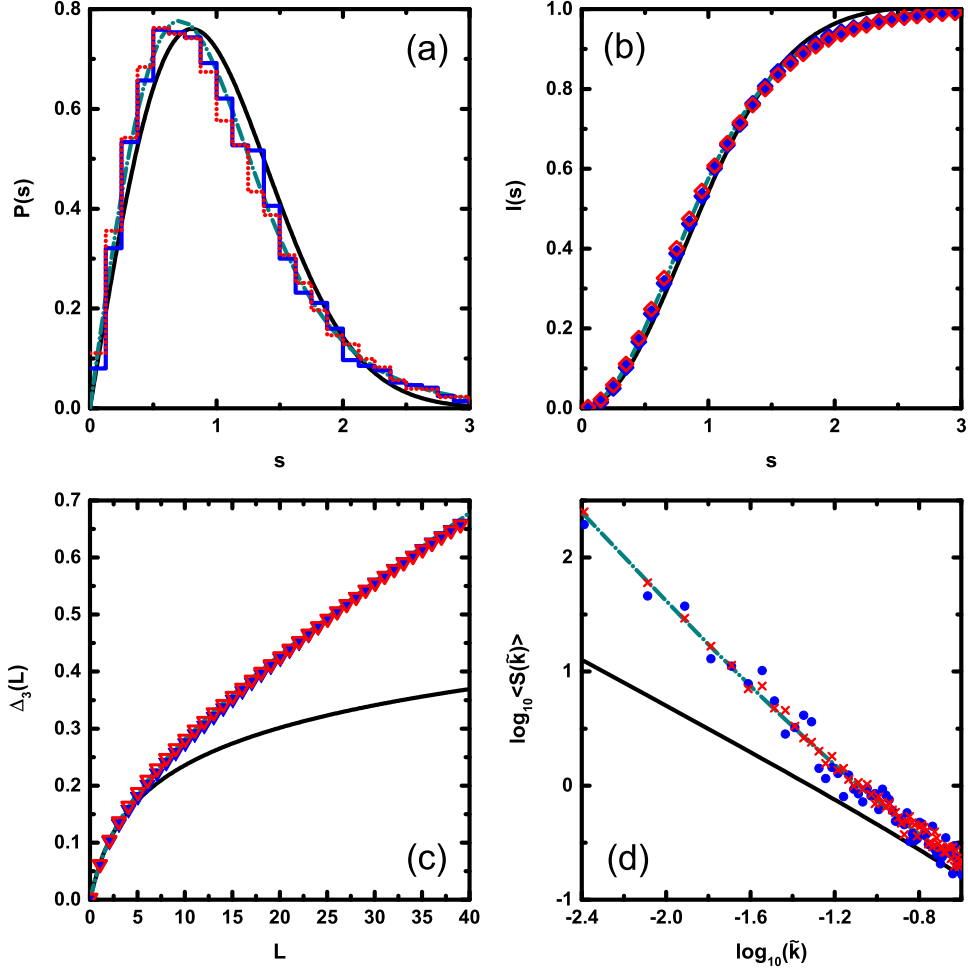}
\caption{
Spectral properties of the rescaled resonance frequencies of the 3D microwave cavity and the power spectrum of level fluctuations. Panel (a) show the nearest-neighbor spacing distribution for the experimental spectra (blue histogram) and for the eigenvalues of random matrices (red doted histogram). Panel (b) show the integrated nearest-neighbor spacing distribution for the experimental spectra (blue diamonds) and for the eigenvalues of random matrices (red empty diamonds). Panel (c) show the spectral rigidity of the spectrum for the experimental spectra (blue triangles) and for the eigenvalues of random matrices (red empty triangles). Panel (d) show the average power spectrum of level fluctuations for the experimental spectra (blue circles) and for the eigenvalues of random matrices (red crosses). The GOE predictions $(\varphi=1)$ and the theoretical predictions evaluated for the fraction of observed levels $(\varphi=0.85)$  are  shown in all four panels by black solid and dark cyan dash-dot lines, respectively.}
\label{Fig4}
\end{figure}

\begin{figure}[h!]
\includegraphics[width=0.9\linewidth]{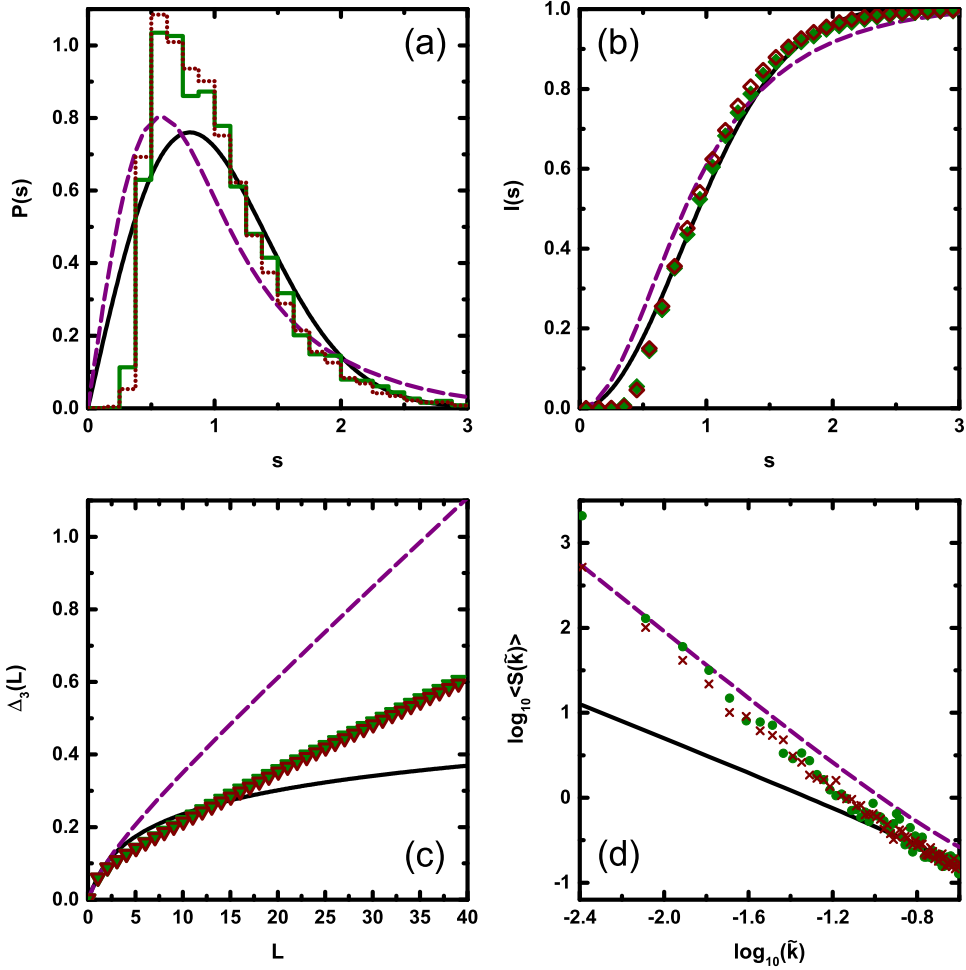}
\caption{
Spectral properties of the rescaled eigenvalues of the 3D microwave cavity and the power spectrum of level fluctuations calculated for the modified experimental spectra from which "the closest" (see the text for details)  resonances were removed to get the fraction of observed levels $\varphi=0.65$. Panels (a)--(d) show the nearest-neighbor spacing distribution (dark green histogram), the integrated nearest-neighbor spacing distribution (dark green diamonds), the spectral rigidity of the spectrum (dark green triangles), and the average power spectrum of level fluctuations (dark green circles), respectively, for modified experimental spectra.  The results obtained from RMT calculations in panels (a)--(c) are marked by dashed line of vine colors, and in panel (d) by vine crosses. The GOE predictions $(\varphi=1)$ and the missing-levels statistics predictions calculated for the fraction of observed levels $\varphi=0.65$ are shown in all panels by full black and purple broken lines,respectively.
}
\label{Fig5}
\end{figure}

\end{document}